\documentclass[pre,aps,epsf,amsfonts,floats,twocolumn,amssymb,amsmath,groupedaddress,showpacs,floatfix,nofootinbib]{revtex4}

\usepackage{graphics}
\usepackage{graphicx}
\usepackage{bm}
\usepackage{amsmath}
\usepackage{amsfonts}
\usepackage{amssymb}
\usepackage{latexsym}
\usepackage{color}

\definecolor{darkred}{rgb}{.8,0,0}

\definecolor{darkblue}{rgb}{0,0,.8}

\newcommand{\be}{\begin{equation}}\newcommand{\ee}{\end{equation}}
\newcommand{\bea}{\begin{eqnarray}}\newcommand{\eea}{\end{eqnarray}}
\newcommand{\brr}{\begin{array}}\newcommand{\err}{\end{array}}
\newcommand{\bit}{\begin{itemize}}\newcommand{\eit}{\end{itemize}}
\newcommand{\ben}{\begin{enumerate}}\newcommand{\een}{\end{enumerate}}

\def\1{{_{1}}}\def\2{{_{2}}}

\begin{document}

\title{Theory of warm ionized gases: Equation of state and kinetic Schottky anomaly}

\author{A. Capolupo${}^{1,2}$}
\author{S. M. Giampaolo${}^{3}$}
\author{F. Illuminati${}^{1,2,4}$}
\affiliation{${}^{1}$ Dipartimento di Ingegneria Industriale, Universit\`a degli Studi di Salerno, Via Giovanni Paolo II, I-84084 Fisciano (SA), Italy}
\affiliation{${}^{2}$ Istituto Nazionale di Fisica Nucleare, Sezione di Napoli, Gruppo collegato di Salerno, I-84084 Fisciano (SA), Italy}
\affiliation{${}^{3}$ University of Vienna, Faculty of Physics, Boltzmanngasse 5, 1090 Vienna, Austria}
\affiliation{${}^{4}$ CNISM - Consorzio Nazionale Interuniversitario per le Scienze Fisiche della Materia, Unit\`a di Salerno, I-84084 Fisciano (SA), Italy}

\pacs{05.70.Ce,64.10.+h,65.40.Ba}

\begin{abstract}
Based on accurate Lennard-Jones type interaction potentials, we derive a closed set of state equations for the description of warm atomic gases
in the presence of ionization processes. The specific heat is predicted to exhibit peaks in correspondence to single and multiple ionizations.
Such kinetic analogue in atomic gases of the Schottky anomaly in solids is enhanced at intermediate and low atomic densities. The case of
adiabatic compression of noble gases is analyzed in detail and the implications on sonoluminescence are discussed. In particular, the predicted plasma electron density in a sonoluminescent bubble turns out to be in good agreement with the value measured in recent experiments.
\end{abstract}

\date{October 6, 2013}

\maketitle

\section{Introduction}

The recent progresses in stellar astrophysics~\cite{Guillot}--\cite{Van Horn}, in experiments on inertial confinement in
plasmas~\cite{Koenig}--\cite{Key}, and in plasma diagnostics have triggered renewed interest
in the determination of accurate state equations for mixed systems of atoms and ions in ionized gases~\cite{Hummer,Rogers,Cardona}. At the same
time, much recent effort has been devoted to investigating the role of plasmas of ionized atomic gases in sonoluminescence
(SL)~\cite{Barber1}--\cite{Flanni}. Sonoluminescence consists in the emission of short flashes of light from collapsing bubbles in a liquid
excited by ultrasonic waves. The light-emitting bubbles contain mainly noble gas atoms plus small quantities of water vapor and in the last part
of the collapse they can reach densities comparable to those of solids. Although the process is extremely fast, the high particle densities and the small bubble sizes allow to assume thermodynamic equilibrium during the entire bubble cycle~\cite{Brenner}. Thus the effect due to charged particle swarms~\cite{Sarchi} can be neglected in the analysis of sonoluminescence. In particular, a well-defined temperature, i.e., local thermodynamic equilibrium must hold even over the short time scales of the bubble dynamics at
collapse. This seems doubtful at first sight. However, the immense particle densities ($n \sim 10^{28} m^{-3}$) and high temperatures ($T \sim 10^4 K$) at bubble collapse create an environment in which collisions between particles are very frequent; therefore, estimated collision times may be well below a picosecond, so that local thermodynamic equilibrium is still well obeyed during single-bubble sonoluminescence light emission~\cite{Brenner}. Moreover, according to recent results~\cite{Flanni}, the conditions inside the
bubbles may be very extreme, the temperature may far exceed those estimated from the emitted light, and the plasma electron density can reach
the same order of magnitude as that of densities created in laser-driven fusion experiments. A detailed understanding of the noble-gas
thermodynamics is thus crucial in order to explain the experimental results and uncover the basic mechanisms of sonoluminescence.

In the present work we analyze the thermodynamics of atomic gases for different regimes of densities and temperatures characteristic of warm
plasmas, and consider applications to the case of noble gases by studying regimes that can be typical of the last stage of collapsing bubbles in
sonoluminescence, i.e. the stage of light emission. Moreover, we study low density regimes which are characteristic of astrophysical objects like stellar atmospheres, interstellar gas, and solar wind~\cite{solar1}.

We derive a set of state equations based on accurate microscopic interaction potentials of the Lennard-Jones type and we predict the presence of
anomalous peaks in the specific heats. These peaks are explained in terms of an extension of the Schottky anomaly observed in solids
at very low temperature~\cite{Tari}.

The peaks appear in correspondence to ionization processes as the temperature is increased. This implies that the different degrees of ionization
play in plasma physics a role similar to the one played by the electron energy levels in the solid state case, the fundamental
difference lying in the fact that the predicted nontrivial contribution to the specific heat in ionized gases is provided by the unbounded
nature of the atomic kinetic energy. The existence of such Schottky anomalies bears important consequences on the collective heating of atomic
gases. In particular, collective heating is strongly suppressed when the specific heats undergo a sharp increase in correspondence of atomic
ionizations. We will discuss in detail the case of noble gases, which is especially relevant for sonoluminescence and various interstellar
phenomena. Considering the late stage of adiabatic compression, we determine self-consistently the number density of plasma electrons and find
good agreement with recently reported experimental results~\cite{Flanni}.

The paper is organized as follows. In Section II, based on accurate microscopic considerations, we define a set of state equations carefully gauged for the investigation of warm plasmas, and we analyze the ultra-low density and the high density plasma regimes. In Section III we consider adiabatic transformations and discuss the kinetic analogue in plasmas of the Schottky anomaly in solids. We then compare the results obtained by using the state equations introduced in Section II with the most reliable existing experimental data on sonoluminescence. Finally, Section IV is devoted to a discussion of the results and future research directions.

\section{Equation of state}

In this section we introduce a new set of state equations for warm ionized gas and we study the plasma regimes at low and high particle densities. We analyze situations comprising mixtures of neutral atoms, ions and free electrons, and we take into account the kinetic, interaction, and ionization contributions. For such physical systems the energy density of the atomic ensemble can be expressed as
\begin{equation}
\label{energy}
 E=\frac{3}{2} K_b T \sum_{i=-1}^{N_{A}} n_i+\sum_{i=-1}^{N_{A}} n_i (E_{i}+h_i) \; ,
\end{equation}
where the index $i$ classifies the particles present in the systems (electrons, atoms, and ions of different charge multiplicity) by their charge
expressed in units of $|e|$. Explicitly, $i=-1$ for free electrons, $i=0$ for neutral atom and $i\in [1,N_{A}]$ for ions of different charge,
ranging up to the maximum allowed value corresponding to the atomic number $N_{A}$. The number density of the $i-$th particle species is denoted
by $n_i$, while $E_i$ is the energy needed to ionize an atom $i$ times, with \mbox{$E_{-1}=E_{0}=0$}. Finally, $h_i$ is the average
potential energy density of the $i$-th particle species due to the interaction with the other particles of the system.
Taking the virial expansion up to second order (higher orders may be considered when necessary), $h_i$ takes the form
\begin{equation}
 \label{fieldsenergy}
  h_{i}\,= \sum_{j=-1}^{N_A} 2 \pi n_j  \int_{0}^\infty  r^{2} U_{i,j}(r) \exp\left(-\frac{U_{i,j}(r)}{K_b T}\right) dr \; ,
\end{equation}
where $U_{i,j}(r)$ is the interaction energy between the $i$-th and the $j$-th particles at distance $r$.
We have assumed throughout that the density $n_j$ is independent on $r$ and is not affected by the presence of the $i$-type of particles.

The atom and ion densities $n_i$ are not fixed {\em a priori}, but are determined by the underlying quantum dynamics through the
self-consistent coupling of Eq.~(\ref{fieldsenergy}) with the Saha equations~\cite{Zeldovic}:
\begin{equation}
\label{Saha}
\frac{n_{i+1} \, n_{-1}}{n_{i}} = \left(\frac{m_e K_b T}{2 \pi \hbar^2}\right)^\frac{3}{2} \frac{2 g_{i+1}}{g_{i}}
\frac{e^{-\frac{E_{i+1}+h_{i+1}+h_{-1}}{K_b T}}}{e^{-\frac{E_{i}+h_{i}}{K_b T}}} \; .
\end{equation}
Here $m_e$ is the electron mass and $g_{i}$ is the ground-state degeneracy of an atom ionized $i$-th times. Eq.~(\ref{Saha}) is a set of
$N_A$ equations in $N_A + 2$ variables represented by the densities $n_i$. The set can be closed taking into account that the electronic
density is related to the ion density by $n_{-1}=\sum_{i=1}^{N_A}i\, n_i$ and that $n_0$ is related to the total atomic density $n$  by
$n_0=n-\sum_{i=1}^{N_A} n_i$.

In agreement with Eq.~(\ref{energy}), the pressure $P$ needs to be written as the sum of a kinetic term and a term associated to the interaction
of every particle with the surrounding ones,
\begin{equation}
\label{Pressure}
P=K_b T\sum_{i=-1}^{N_A} n_i - 2 \pi \sum_{i=-1}^{N_A} n_i h^{int}_{i} \, ,
\end{equation}
where the interaction term $h^{int}_{i}$, considering the virial expansion up to second order, reads
\begin{equation}
\label{fieldspressure}
h^{int}_{i}= \sum_{j=-1}^{N_A} n_j \int_{0}^\infty  r^{3} \frac{dU_{i,j}(r)}{dr} \exp\left(-\frac{U_{i,j}(r)}{K_b T}\right) dr  \, .
\end{equation}
By means of Eq.~(\ref{energy}) we can compute the specific heat at constant volume, $c_v=\left(\frac{\partial E}{\partial T}\right)_V$.
Eq.~(\ref{Pressure}) can be used to obtain the specific heat at constant pressure by using the relation
\mbox{$c_p-c_v=-T \left(\frac{\partial P}{\partial V}  \right)_{T}^{-1} \left( \frac{\partial P}{\partial T}  \right)_{V}^{2}$}.
Eqs.(\ref{energy})--(\ref{fieldspressure}) identify a complete set of state equations for a gas in which the degree of ionization is
determined self-consistently from thermal equilibrium.

\subsection{Non interacting systems: warm plasmas at low density}

We consider first the situation of warm plasmas at ultra-low densities, such that the interaction effects can be neglected, i.e. $U_{i,j}=0 \,, \, \forall i,j$.
Such a regime is characteristic of stellar atmospheres, interstellar gas, and solar winds, which are extensively analyzed in astrophysical research~\cite{solar1}.

\begin{figure}
\includegraphics[width=7.5cm]{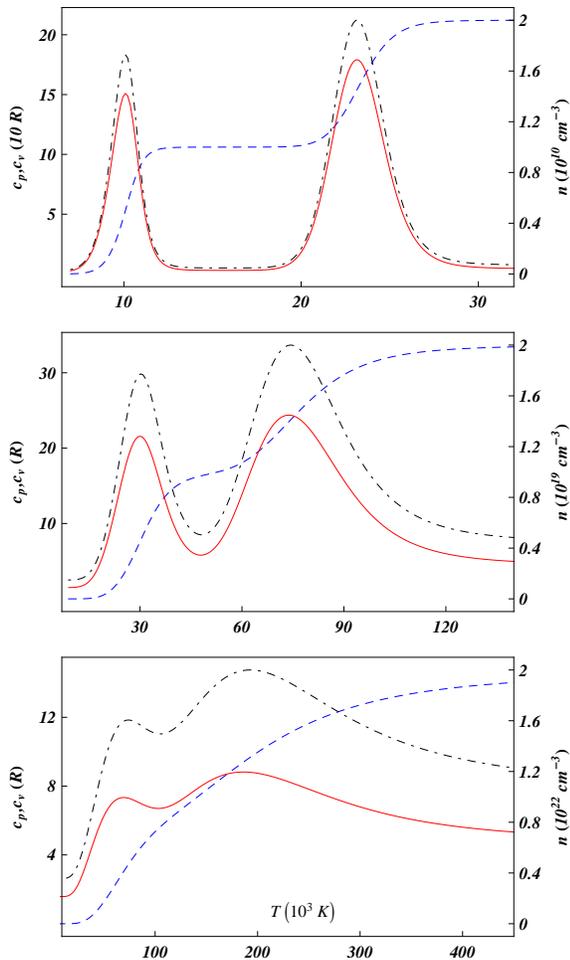}
\caption{The molar specific heats at constant volume $c_v$ (Red solid line), and at constant pressure $c_p$ (black dot-dashed line) as
functions of the temperature $T$, compared with the electron density $n_{-1}$ (blue dashed line) for an ensemble of He atoms at three different
densities $n$. Top panel: $n= 10^{10} cm^{-3}$; central panel: $n = 10^{19} cm^{-3}$; bottom panel: $n = 10^{22} cm^{-3}$. The specific heats
are expressed in units of the universal gas constant $R$.} \label{cvcpideal}
\end{figure}
\begin{figure}
\includegraphics[width=7.5cm]{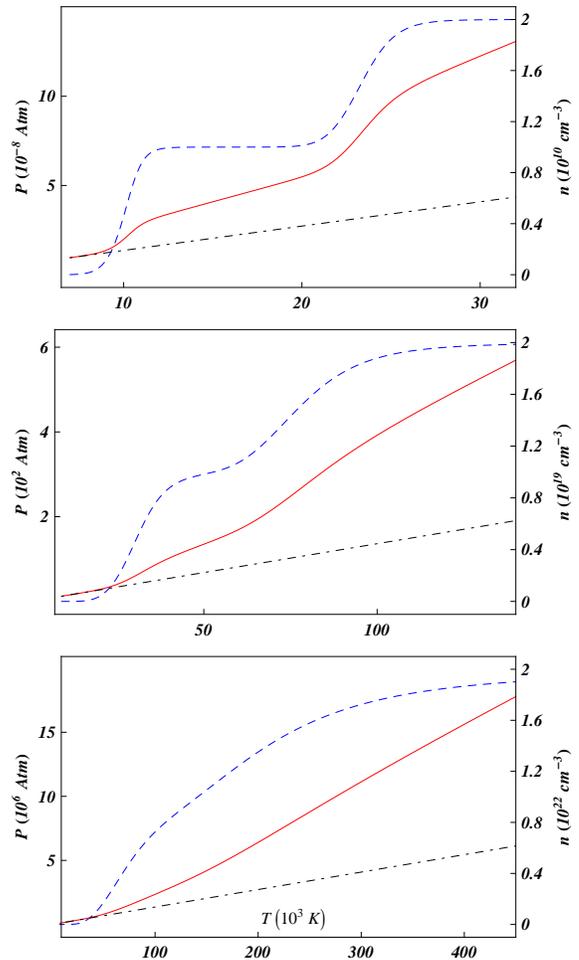}
\caption{The pressure $P$ (red line) as function of the temperature $T$, compared with the electron density $n_{-1}$ (blue dashed line) and
with the pressure obtained from the ideal gas law neglecting ionization processes (black dot-dashed line), for an ensemble of He atoms at
the same three values of the density reported in Fig.~\ref{cvcpideal}.} \label{pressureideal}
\end{figure}
For this regime, the main results, obtained solving the set of Eqs.(\ref{energy})--(\ref{fieldspressure}), are illustrated in
Figs.~\ref{cvcpideal} and \ref{pressureideal} for a system of He atoms. We show the
behavior of the pressure $P$, the specific heats $c_v$ and $c_p$, and the density of free electrons $n_{-1}$ as functions of the temperature
$T$ at different values of the total density. The top panel of both figures refers to the instance of $n=10^{10} cm^{-3}$, the typical density in
the active galactic nucleus~\cite{solar1} and about ten orders of magnitude less than the corresponding value at room temperature on earth's
surface. From Fig.~\ref{cvcpideal} we observe that both specific heats exhibit two peaks in exact correspondence,
respectively, with the processes of first and second ionization, as identified by the associated jumps in the free-electron density $n_{-1}$.
Correspondingly in Fig.~\ref{pressureideal} the pressure shows a step-wise behavior associated to
the increase in the number of free electrons released in the ionization process.

Going to regimes of much higher densities, as reported in the central and bottom panels of Figs.~\ref{cvcpideal} and \ref{pressureideal},
we observe the progressive smoothing out of the peaks in the specific heats, as well as the broadening of the step-wise behavior of the density
$n_{-1}$ and of the pressure $P$. This is due to the fact that as the density increases,
the respective ranges of temperature in which the processes of first and second ionization occur start to merge due to an increasing recombination probability between ions and free electrons. As a consequence, one observes an overall increase in the effective ionization temperature of the system.

The onset of the peaks occurs as the temperature increases and the corresponding thermal energy approaches the difference between the
atomic energy levels.
In this situation the different ionization states become significantly populated and hence there is an enormous increase in
the entropy of the system due to the uncertainty on the degree of ionization. This implies the presence of peaks in the profiles of the specific heats.

At variance with the Schottky anomaly occurring in solid state physics for systems
with a finite number of energy levels (bound states), in the case of non interacting plasmas the rate of change of the entropy is due to the kinetic energy of the ionized atoms, resulting in multiple peaks of the specific heats, one for each successive degree of ionization. Therefore this a kind of \emph{kinetic} Shottky anomaly, in principle with an unbounded number of peaks in the specific heats.

\subsection{Interacting systems: warm dense plasma}

We now include the interactions between different particles in our investigation by introducing two-point interatomic potentials $U_{i,j}$ which
can describe with reasonable accuracy the microscopic physics of plasmas in the intermediate regimes of temperature and density, i.e. warm dense
plasmas. Here and in the following we will model the interaction between two neutral atoms of a noble gas by Lennard-Jones type potentials:
\mbox{$ U_{0,0}(r) = 4\varepsilon  \left[ \left( \frac{\sigma}{r} \right)^{12} - \left( \frac{\sigma}{r} \right)^6\right]$},
in which $\varepsilon$ stands for the depth of the potential well and $\sigma $ is the finite distance at which the inter-particle potential
vanishes.

The Lennard-Jones potential that we propose to adopt is well justified as it amounts to the sum of a short-range Pauli repulsion due to
the partial overlap of the electronic clouds, and a long-range dipole-dipole attraction between the two neutral atoms. We need also to include
the interaction energy between a neutral atom and an ion, which can be described by adding to $U_{0,0}$ the charge-dipole attractions
\mbox{$ U_{0,i}(r) = U_{0,0}(r) - i^{2} \alpha E_{l}^{2}(r)$}, where $\alpha$ is the polarizability of the neutral atom and $E_{l} (r)$ is the
electric field associated to the electrostatic potential $W_{l}(r)$ that solves the corresponding Debye-H\"{u}ckel equation~\cite{Zeldovic}.
Following the same idea, we describe  ion-ion interactions by further adding the charge-dipole interactions and the electrostatic ion-ion
repulsion: \mbox{$ U_{i,j}(r) = U_{0,0}(r) - (i^{2}+j^{2}) \alpha E_{l}^{2}(r)+i j W_{l}(r)$}.

The interactions involving one or more free electrons can be described by considering any free electron as a cloud similar to the one surrounding
an Hydrogen atom, but with a radius equal to the thermal De Broglie wavelength $\lambda$ and a vanishing polarizability. With these assumptions,
free electrons can be treated  as ions and their interactions similarly to the ones between positive, massive ions. We have only to replace, in
the Lennard Jones term, $\sigma$ with $\frac{\sigma+\lambda}{2}$ and $\varepsilon$ with $\sqrt{\varepsilon \, \varepsilon_H}$ (being
$\varepsilon_H$ the depth of the potential well of the Lennard-Jones potentials between two Hydrogen atoms) in the case of an interaction
between a free electron and an ion, and to replace  $\sigma$ with $\lambda$ and $\varepsilon$ with $\varepsilon_H$ in the $U_{-1,-1}(r)$ term.

\begin{figure}
\includegraphics[width=7.5cm]{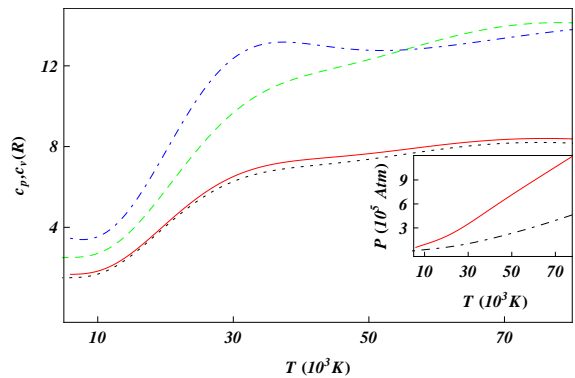}
\caption{Behavior of the specific heats as functions of the temperature, with and without interactions, for a molar weight of Xenon. Main plot:
$c_v$ for the non interacting case (dotted black line), $c_v$ for the interacting case (solid red line), $c_p$ (green dashed line) for the non
interacting case, and $c_p$ (blue dot-dashed line) in the interacting case. Inset: behavior of the pressure as a function of the temperature,
with and without interactions. Black dot-dashed line: pressure for the non interacting case. Red solid line: pressure for the interacting case.
In all cases the density is fixed at $n=10^{22} cm^{-3}$, and only first and second ionizations are included.} \label{correction}
\end{figure}

In order to compare the cases with and without interactions, we consider here and in the following a system of Xenon atoms at high density.
The choice of Xenon is motivated by the fact that this element is characterized by a lower ionization energy than that of
Helium~\cite{Emsley}. Hence, one may reasonably expect that, especially at high density, the corrections due to interactions will be
sizable and clearly detectable. In our further treatment we will neglect the dependence of $\sigma$ and $\varepsilon$ on the degree of
ionization. This approximation does not affect results in the case in which one considers only the first few ionization levels.
Therefore, in the analysis summarized in Fig.~\ref{correction} we take into account only the processes of first and second ionization in
Xenon at densities of the order $n=10^{22} cm^{-3}$.

Fig.~\ref{correction} shows that particle interactions do not modify significantly the Shottky anomaly in $c_v$, even at high densities.
Indeed, the fields $h_i$ in Eq.~(\ref{fieldsenergy}) are about one order of magnitude lower than the ionization energy and therefore their effect
on the solution of the Saha Equation Eq.~(\ref{Saha}) is negligible. On the other hand, the behaviors of $P$ and of $c_p$ are strongly
influenced by the presence of interactions, as the latter introduce relevant Newtonian effects in the conservative forces and in the ensuing
associated pressures, whose net effect is the strong suppression of the Schottky-like type of behavior.

Notice that in our treatment we neglected the electronic excitations of warm plasmas. Such excitations also consume energy and contribute to the increase in the specific heat. However, their contributions are much less important than the ones due to ionization, and therefore the
behaviors of $c_p$ and $c_v$ presented in Figs.(1) and (3) can be considered good approximations of the real plasma behavior.

\section{Schottky anomaly and adiabatic compression}

Next, we consider the influence of the Schottky anomaly discussed above on the temperature achievable in a process of adiabatic compression. We focus on this particular thermodynamic transformation because it plays a fundamental role in different atomic phenomena of great recent interest, such as sonoluminescence. Indeed, in sonoluminescence the (mostly noble) gas inside a bubble follows an isothermal transformation during the expansion of the bubble and an adiabatic compression during the bubble's collapse, the latter being the
characterizing phase for the emission of light~\cite{Brenner}.

In principle, Schottky anomalies in correspondence of ionization processes should have a sizeable effect on the collective heating and on the temperature achievable in an adiabatic compression. On the other hand, these types of transformations for gases in intermediate regimes of density and temperature are typically
described by the equations of state of ideal gases supplemented by their van der Waals refinements to account for corrections due to interactions. In either case, ionization processes cannot be handled properly, and Schottky-like behaviors cannot be predicted. We will thus compare the description of adiabatic compression obtained using the ideal-gas equation of state, the van der Waals one, and our self-consistent set based on the full microscopic description of atoms and ions.

To set the framework, in the following we consider a quasi-static adiabatic compression of a sphere containing one mole of atoms which is initially in thermal equilibrium under standard conditions at room temperature: $P=1 atm$, $T=300 K$. An adiabatic compression is performed until the density in the sphere becomes comparable with the one present in a sonoluminescent bubble during the last stage of its collapse. We investigate the predictions on the behavior of the temperature inside the bubble after the compression, according to the standard state equations (ideal or van der Waals) and to our set of state equations combining the description of the atomic interactions and the ionization processes with Lennard-Jones type potentials in a unified framework.
\begin{figure}
\includegraphics[width=7.5cm]{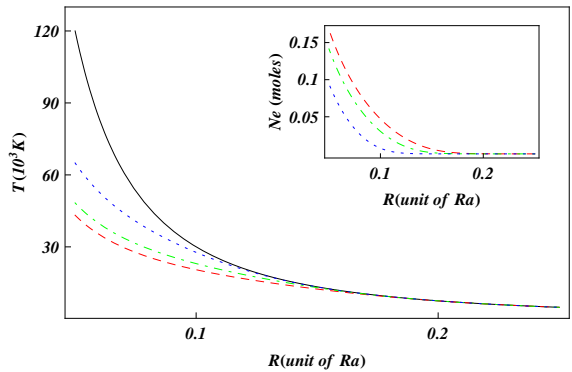}
\caption{Main plot: behavior of the temperature inside a bubble containing a mole of atomic gas as a function of the radius of the bubble
(expressed in unit of the ambient radius $R_a$) after an adiabatic compression. The prediction of the ideal-gas equation of state
(black solid line) is compared with those obtained from our set of state equations for non interacting Helium (blue dotted line), non
interacting Argon (green dot-dashed line) and non interacting Xenon (red dashed line). Inset: total number of free electrons, as a function of
the bubble radius, in the case of an adiabatic compression for the same atomic species considered in the main plot.} \label{Adiabaticcompression}
\end{figure}

The results of our comparison are reported in Fig.~\ref{Adiabaticcompression}. Notice that when the radius of the sphere $R$ is sufficiently large, the solutions obtained modeling the atomic mole as an ideal gas are indistinguishable from the microscopic model Eqs.(\ref{energy})--(\ref{fieldspressure}), as it should be. However, when $R$ becomes comparable to or smaller than $0.2\; R_a$, where $R_a \simeq 18.04 \; cm$ denotes the radius of the sphere at standard room temperature conditions, the central temperatures predicted by state equations Eqs.(\ref{energy})--(\ref{fieldspressure}) are considerably lower than the ones obtained
for the ideal gas. The lowering of the temperature is due to the fact that a part of the energy provided to the system by the compression is used up by some of the atoms into ionization processes. Xenon is the gas in which this effect is more relevant because its ionization energies are lower than the ones of Argon and Helium. Even if the number of free electrons is low compared to the number of non-ionized atoms, and hence the system is only at the onset of the Schottky anomaly, the effect is so strong to halve the temperature as, e.g., in the Helium case at $R=0.05 R_a$.

The correction is even stronger if we compare our results with the temperatures predicted using the van der Waals state equation. Indeed, the latter induces an unphysical divergence in the temperature when the volume of the bubble approaches the co-volume of the atomic species involved.

Finally, from the inset in Fig.~\ref{Adiabaticcompression} we observe that for densities comparable with the ones typical in sonoluminescent bubbles, the electronic density obtained with the set of state equations (1)-(5) is in good agreement with the most accurate recent experimental results on sonoluminescence~\cite{Flanni}, according to which the plasma electron density in the bubble exceeds $10^{21} cm^{-3}$. This result indicates that the plasma state equations (1)-(5) that we have introduced can be useful for the description of the plasma formed inside the sonoluminescent bubbles, whose presence has already been revealed experimentally~\cite{Brenner,Suslick2,Suslick5,Xu,Flanni}. Such equations can improve the current theories based on thermal ionized plasma (see detailed reviews as~\cite{Brenner,Suslick5}).

In the thermal emission a large variety of different processes are important. Since temperatures are increasing from several hundreds to many thousands Kelvin during collapse, the relevant processes can include molecular recombination, collision-induced emission, molecular emission, excimers, atomic recombination, radiative attachments of ions, neutral and ion Bremsstrahlung, and emission from confined electrons in voids. Which theoretical scheme applies depends then on the accurate measurements and calculations of the temperature inside the bubble.

By analyzing the relative intensities of the atomic emission lines, temperatures of up to $1.5 \times 10^{4}K$ have been estimated during single-bubble cavitation. However, these temperatures may not reflect the core temperature within the collapsing bubble. Plasma formation may lead to an optically opaque region within the bubble that is characterized by much higher temperatures and pressures~\cite{Hiller}. The observed temperature may well only reveal the conditions at the outer shell of
the inner opaque core~\cite{Suslick2,Suslick5,Xu,Flanni}. Eqs.(1)-(5) can then allow the study of ranges of temperature which could be characteristic of the inner hot core and therefore they could allow a more precise investigation and a better understanding of the mechanisms ruling the sonoluminescence phenomenon.

\section{Discussion and outlook}

We have introduced a set of state equations for warm ionized gases, characterized by the fact that the interatomic, atom-ion, and interionic interactions are modeled by accurate microscopic potentials of the Lennard-Jones type, and the populations in the sequential ionization processes are not fixed arbitrarily {\em a priori} but are obtained self-consistently using the Saha equation.

We have analyzed in detail the case of noble gases, because of their importance in many physical phenomena such as sonoluminescence. However, our work can be extended straightforwardly also to different kinds of atoms, molecules, and mixtures of them by allowing for chemically bounded elements. Using this new approach we have shown that the noble gas atoms at low densities exhibit a series of anomalies (fast increase
followed by a fast decrease) in their specific heats, that represent the kinetic analogous for continuous energy spectra of the Schottky anomalies observed in the specific heats of solid-state systems with discrete and finite energy spectra.

Moreover, we have shown that pressure and density of the free electrons exhibit a step-wise behavior in correspondence of the different ionization processes. The anomalies in
specific heats become less evident at high densities and disappear when all the atoms are fully ionized.
We have also shown that interactions affect relevantly both the pressure and the specific heat at fixed pressure $c_p$ while, on the contrary, the specific heat at fixed volume $c_v$ and the electron density $n_{-1}$ remain essentially unaffected.

This novel type of Schottky anomalies predicted for the specific heats of noble gas atoms can play a very relevant role in many phenomena of actual interest, for instance sonoluminescence, in which collapsing bubbles close to their minimum radius are almost completely filled with noble gas atoms, reach temperatures of few tens of thousands of Kelvin, and densities of the order of the ones typical of solids~\cite{Brenner,Lohse}. In such phenomena, during the last stage of the collapse phase, the temperature and the radial speed of the bubble depend crucially on $c_p$ and $c_v$, so that their increase in correspondence of a Schottky anomaly can influence the dynamics of the bubble quite strongly.

In the present paper we have not directly considered the problem of the full dynamics of sonoluminescence, which will be analyzed in detail elsewhere, but we have only applied our general findings on the thermodynamics of warm ionized gases to the simple problem of the adiabatic compression of a sphere that reaches a final density comparable with the one of sonoluminescence during the emission of light, and
we have predicted values of the electronic density for the gas in the sphere that are in very good agreement with the most accurate experimental results currently available~\cite{Flanni}. In this respect, the present study may also be useful for the analysis of stellar atmospheres, interstellar gases, and solar winds.

Prompted by these encouraging results we are currently exploring the possibility of more refined investigations. These will need to address a thorough treatment of the sonoluminescent bubble dynamics that will need to include, beyond the set of state equations that we have introduced in the present work, dynamical equations of the Rayleigh--Plesset type. Our methods can also be generalized to include the layered analysis of the sonoluminescent bubble's central region, in order to obtain first-principle predictions on the temperature of the hot core in collapsing bubbles, a crucial problem that remains so far unsolved both theoretically and experimentally. These and related issues will be the subject of forthcoming work.

\section*{Acknowledgment}
We acknowledge financial support from the European Commission of the European Union under the FP7 STREP Project iQIT (integrated Quantum Information Technologies), Grant Agreement n. 270843.

\end{document}